\documentstyle[prb,aps]{revtex}
\begin{document}
\mediumtext
\draft
\title{
General exact solution of the Einstein--Dirac equations \\ with the
cosmological constant in homogeneous space}

\author{
V.\ A.\ Zhelnorovich\thanks{Electronic mail: zhelnor@inmech.msu.su}}

\address{
Institute of Mechanics,
Moscow state university,
Michurinsky pr., 1, Moscow 119899, Russia.}
\maketitle

\begin{abstract}
The general exact solution of the Einstein--Dirac
equations with cosmological constant in the homogeneous
Riemannian space of the Bianchi 1 type is obtained.
\end{abstract}


\section{INTRODUCTION}
While integrating the Einstein--Dirac equations, one has to face two
difficulties.

	The first one is a purely technical difficulty  related to the
Einstein--Dirac equations being a complicated system of nonlinear partial
differential equations of the second order with 24 unknown functions.
Some particular exact solutions \cite{1,2,3,4,5,6} of the Einstein-Dirac
equations in homogeneous spaces  used to be obtained before only
for the diagonal metrics of Riemannian space.

    The second difficulty is of a fundamental nature and related to the
spinor field in Riemannian space being definable only in some nonholonomic
orthonormal bases (tetrads) to be set. In other words,  a gauge of the
tetrads is necessary. Lots of such gauges are known, and different authors
proposed different gauges. Taken together these gauges either are
noninvariant with respect to  transformation of variables in the observer's
coordinate system or  are written in the form of differential equations,
which results in a complication of the initial system of equations.

 	Physically,  all gauges are equivalent, since the Einstein-Dirac
equations are tetrad--choice invariant. Mathematically, the use of a bad
gauge (i.e. additional equations closing the Einstein-Dirac ones) may
seriously complicate the equations, but the use of a good gauge may
significantly simplify them.

 To the difficulty in a reasonable gauge of tetrads detect is much related
that in  previous papers  \cite{1,2,3,4,5,6} solutions of the
Einstein--Dirac equations have been obtained  only for diagonal metrics (for
such metrics the vectors of the basis of a holonomic coordinate system are
orthogonal, and hence it is possible to make a natural choice of the tetrads
related with an orthogonal holonomic basis of Riemannian space).

   Here we use the tetrad gauge \cite{7} that is algebraic and, at the same
time, is formulated in an invariant way. Using this gauge allows one to
reduce the number of unknown functions in the Einstein--Dirac equations by six
units conserving the invariance of equations with respect to transformation
of variables of the observer's coordinate system.

With the tetrad gauge
being used here, all equations are formulated as the
first-order equations only for two invariants of the spinor field and
Ricci rotation coefficients of the proper bases defined by the spinor field.
The Dirac equations after the tetrad gauge transform into equations for Ricci
rotation coefficients and for invariants of the spinor field, with the Ricci
rotation coefficients entering in these equations linearly and
derivative-free. Therefore in a homogeneous Riemannian space the Dirac
equations prove to close the Einstein equations for the Ricci
rotation coefficients without using additional equations. In this case one
may first integrate the first order equations for the Ricci rotation
coefficients and the spinor field invariants, and then integrate the first
order equations for scale factors.

With these two circumstances --- a reduction of the number of unknown
functions by six units and a possibility of integrating the second
order equations in two steps --- considerable
simplification  of the Einstein--Dirac equations is related, which makes it
possible to obtain new exact solutions of these equations.

\section{
Spinor fields in the four-dimensional Riemannian space
  }

Let $V$ is the four-dimensional Riemannian space with the metric
signature $(+,+,+,-)$, referred to a system of coordinate with the
variables $x^i$ and the holonomic vector basis ${\cal J}_i$,
$i=1$, 2, 3, 4. The metric tensor of the space $V$ is determined
in basis ${\cal J}_i$ by covariant components $g_{ij} $;
the connection is determined by the Christoffel symbols $\Gamma _{ij}^s$.
Let us introduce in the space $V$ a smooth field of the orthonormal
bases (tetrads) ${\bf e}_a(x^i)$ ($a=1$, 2, 3, 4), by the relations
$$
{\bf e}_a=h^i{}_a{\cal J}_i, \qquad
{\cal J}_i=h_i{}^a{\bf e}_a,
\eqno (1.1)
$$
where $h_i{}^a$, $h^i{}_a$ are the scale factors. We designate the indices
of tensor components in the basis ${\cal J}_i$ by Latin letters $i$,
$j$, $k$, ... ; the indices of tensor components in the orthonormal basis
${\bf e}_a$ will be designated by the first letters of the Latin
alphabet $a$, $b$, $c$, $d$, $e$, $f$.

The differential of the orthonormal basis vectors ${\bf e}_a(x^i)$
is defined by the Ricci rotation coefficients
$$
d{\bf e}_a =\Delta _{i,a}{}^b{\bf e}_bdx^i,
\eqno(1.2)
$$
which are expressed via scale factors
$$
\Delta _{i,ac}=\frac 12\bigl[
h^j{}_c\bigl( \partial _ih_{ja}-\partial _jh_{ia}\bigr)
-h^j{}_a\bigl( \partial _ih_{jc}-\partial _jh_{ic}\bigr)
+h_i{}^bh^j{}_ah^s{}_c\bigl(\partial _jh_{sb}
-\partial _sh_{jb}\bigr)\bigr].
\eqno(1.3)
$$

Here $ \partial _i =\partial /\partial x^i $ is the symbol of a partial
derivative with respect to the variable $x^i $.

Let us determine in the Riemannian space $V$ a spinor field of the first
rank $ {\bf\psi }(x^i) $, given by contravariant components
$ \psi ^A (x^i) $ ($A=1$, 2, 3, 4) in the orthonormal bases
$ {\bf e}_a (x^i) $.
The spinor indeces of the components of any spin-tensor are lowered or
raised by the formulas
$$
\psi ^A=e^{AB}\psi _B, \qquad \psi _A=e _{AB}\psi ^B,
\eqno(1.4)
$$
where $E=\| e_{AB}\|$, $E^{-1}=\|e^{AB}\| $ are the covariant
and contravariant components of the metric spinor, defined
by the equations
$$
\gamma _a^T=-E\gamma _aE^{-1}, \qquad E^T =-E.
\eqno(1.5)
$$

Here $T$ is the transposition symbol; $\gamma _a$ are the
four-dimensional Dirac matrices that on definition satisfy the
equation
$$
\gamma _a\gamma _b +\gamma _b\gamma _a=2g_{ab}I,
\eqno(1.6)
$$
where $I$ is the unit four-dimensional matrix,
$\| g_{ab}\| =\mathop{\rm diag} (1,1,1,-1)$
the covariant components of  metric tensor in
an orthonormal basis ${\bf e}_a$.

It should be noted, that in the equations (1.4) contracting
is always done with respect to the second index of antisymmetric
components of the metric spinor.

Let us define also the conjugate spinor field with the covariant
components $ \psi ^+ =\|\psi ^+_A\|$ by the relation
$\psi ^+=\dot\psi ^T\beta$, where the point above a letter denotes
the complex conjugation; the invariant spinor of second rank
$\beta$ is defined by the equations
$$
\dot\gamma _a^T =-\beta\gamma _a\beta ^{-1}, \qquad
\dot \beta ^T =\beta.
\eqno(1.7)
$$

The four-component spinor field $ \psi $ in general case has two
real invariants $ \rho $, $ \eta $, which can be defined
by the equation
$$
\rho \exp i\eta = \psi ^+\psi +i\psi ^+\gamma ^5\psi,
\eqno(1.8)
$$
where $ \gamma ^5 =\frac 1{24}\varepsilon ^{abcd} \gamma _a\gamma_b
\gamma _c\gamma _d $, $\varepsilon ^{abcd} $ are the components  of the
four-dimensional Levi--Chivita tensor  $\varepsilon ^{1234}=-1$.
In the Riemannian space $V$ via the spinor field $ \psi $ and
conjugate spinor field $\psi ^+$ it
is possible to define the proper orthonormal basis
$\breve {\bf e}_a $ of the spinor field
$$
\breve {\bf e}_1=\pi ^i {\cal J}_i,   \qquad
\breve {\bf e}_2 =\xi ^i {\cal J}_i,  \qquad
\breve {\bf e}_3=\sigma ^i {\cal J}_i,\qquad
\breve {\bf e}_4 =u^i {\cal J}_i.
\eqno(1.9)
$$
The vector components $ \pi ^i $, $ \xi ^i $, $ \sigma ^i $,
$u^i $ in the equations (1.9) are defined by the relations \cite{8,9}
$$
\begin{array}{rcl}
\rho \pi ^i&=&\mathop{\rm Im} (\psi ^TE\gamma ^i\psi ), \\
\rho \xi ^i&=&\mathop{\rm Re} (\psi ^TE\gamma ^i\psi ), \\
\rho \sigma ^i&=&\psi ^+\gamma ^i\gamma ^5\psi,           \\
\rho u^i&=&i\psi ^+\gamma ^i\psi,
\end{array}
\eqno(1.10)
$$
where the invariant $\rho $ is determined by the relation (1.8),
the spin-tensors $\gamma ^i=h^i{}_a\gamma ^a$ satisfy the equation
$$
\gamma ^i\gamma ^j +\gamma ^j\gamma ^i=2g ^{ij} I.
\eqno(1.11)
$$

It is obvious, that the scale factors $ \breve h^i{}_a $,
corresponding to the proper basis
$ \breve {\bf e} _a $, are determined by  matrix
$$
\breve h^i{}_a=\left\|
\begin{array}{cccc}
	  \pi ^1&\xi ^1&\sigma ^1&u^1\\
		 \pi ^2&\xi ^2&\sigma ^2&u^2\\
		 \pi ^3&\xi ^3&\sigma ^3&u^3\\
		 \pi ^4&\xi ^4&\sigma ^4&u^4 \end{array}\right\|.
\eqno(1.12)
$$

If the spin-tensors $ \gamma _a $, $E $, $ \beta $ are determined
by matrices
$$      
\begin{array}{c}  \vspace{2\jot}
 \gamma _1=\left\|\begin{array}{cccc}
                   0&0&0&i\\
                   0&0&i&0\\
                   0&-i&0&0\\
                  -i&0&0&0   \end{array}  \right\|,  \qquad
\gamma _2=\left\| \begin{array}{cccc}
                   0&0&0&1\\
                   0&0&-1&0\\
                   0&-1&0&0\\
                   1&0&0&0 \end{array} \right\|,    \qquad
\gamma _3=\left\|  \begin{array}{cccc}
                   0&0&i&0\\
                   0&0&0&-i\\
                  -i&0&0&0\\
                   0&i&0&0 \end{array}\right\|,  \\  \vspace{2\jot}
\gamma _4=\left\| \begin{array}{cccc}
                   0&0&i&0\\
                   0&0&0&i\\
                   i&0&0&0\\
                   0&i&0&0 \end{array}\right\|,  \qquad
E=\left\| \begin{array}{cccc}
            0&1&0&0\\
 	   -1&0&0&0\\
            0&0&0&-1\\
            0&0&1&0 \end{array}\right\|, \qquad
\beta =\left\| \begin{array}{cccc}
 		   0&0&1&0\\
		   0&0&0&1\\
		   1&0&0&0\\
		   0&1&0&0
		      \end{array}   \right\|,
\end{array}
\eqno(1.13)
$$
then the components of spinor ${\bf\psi} $ calculated in the proper
basis $\breve {\bf e}_a $, are determined by the
invariants $ \rho $, $ \eta $ as follows \cite{10}
$$
\begin{array}{rcl}
\breve\psi^{\, 1}=0,&& \quad
\breve\psi^{\, 2}=
i\sqrt{\frac 12\rho}\,\exp\left(\frac i2\eta\right), \\
\breve\psi^{\, 3}=0,&&\quad
\breve\psi ^{\, 4}=
i\sqrt{\frac 12\rho}\,\exp\left( -\frac i2\eta\right).
\end{array}
\eqno(1.14)
$$

As is well known, the covariant derivative of spinor fields $\psi$,
$\psi ^+$ in the Riemannian space are defined by the relations \cite{11}
$$
\nabla _s\psi =\partial _s\psi
-\frac 14\Delta _{s,ij}\gamma ^i\gamma ^j\psi, \qquad
\nabla _s\psi ^+=\partial _s\psi ^+
+\frac 14\psi ^+\Delta _{s,ij}\gamma ^i\gamma ^j.
\eqno(1.15)
$$

For the covariant derivative of spinor fields $\psi $, $\psi ^+$
the equations are valid
$$
\begin{array}{c}
\nabla _s \psi =\left(
-\displaystyle\frac 14\breve\Delta _{s,ij}\gamma ^i\gamma ^j+
\frac 12I\partial _s\ln\rho -\frac 12\gamma ^5\partial _s\eta\right)\psi, \\
\nabla _s \psi ^+= \psi ^+
\left( \displaystyle\frac 14 \breve\Delta _{s,ij} \gamma ^i\gamma ^j +
\frac 12I\partial _s\ln \rho -\frac 12 \gamma ^5\partial _s\eta \right).
\end{array}
\eqno(1.16)
$$
Here the invariants of spinor field $\rho
$, $\eta $ are defined by the equation (1.8), the Ricci rotation coefficients
$\breve\Delta _{i,jk}=\breve h_j{}^b\breve h_k{}^c\breve\Delta _{i,bc}$
correspond to the proper bases $\breve {\bf e}_a $
of a spinor field and are calculated by the formula
$$
\breve\Delta _{s,ij}=\frac 12\bigl( \pi _i\nabla _s\pi _j
-\pi _j\nabla _s\pi _i+\xi _i\nabla _s\xi _j-\xi _j\nabla _s\xi _i
+\sigma _i\nabla _s\sigma _j
-\sigma _j\nabla _s\sigma _i-u_i\nabla _su_j+u_j\nabla _su_i\bigr).
\eqno(1.17)
$$

The relations (1.16) in the four-dimensional pseudoeuclidean space have
been obtained in \cite{12,13}. The relations (1.16) in a Riemannian space can
be obtained from the corresponding relations in pseudoeuclidean space by
replacement particular derivative to covariant ones. The relations (1.16)
are satisfied identically in view of definitions of the quantities
$\rho $, $\eta $, $\breve\Delta _{s,ij}$.

  \section{
Einstein--Dirac equations with the cosmological constant
  }

Let us consider the set of equations
$$\begin{array}{rcl}
\gamma ^a\nabla _a\psi +m\psi &=&0,    \\
R_{ab}-\displaystyle\frac 12 Rg_{ab}+\lambda g_{ab}&=&\kappa T_{ab}.
\end{array}
\eqno(2.1)
$$

Here $ \psi $ is a four-component spinor field in the four-dimensional
Riemannian space of events $V$, given in the orthonormal basis
$ {\bf e}_a $; $m$, $ \lambda $, $\kappa $ are constants;
$R=g^{ab}R_{ab}$ is the scalar curvature of space $V$,
$R^{ab}$ the components of the Ricci tensor calculated in basis
${\bf e}_a$ and $g^{ab}=\mathop{\rm diag}(1,1,1,-1)$.

The components $T_{ab}$ of the energy--momentum tensor of the spinor
field in basis ${\bf e}_a$ are defined by the relation
$$
T_{ab}=\frac 14 \left( \psi ^+\gamma _a\nabla _b\psi -\nabla _b\psi ^+
\cdot\gamma _a\psi + \psi ^+\gamma _b\nabla _a\psi -\nabla _a\psi ^+
\cdot\gamma _b\psi \right).
\eqno(2.2)
$$

The equations (2.1) are invariant under the arbitrary pseudoorthogonal
transformations of tetrads ${\bf e}_a$, therefore for the closure
of equations (2.1) it is necessary to add the additional equations,
defining tetrads ${\bf e}_a$. Such additional equations usually are
called as the gauge conditions. Further on we accept the gauge conditions
${\bf e}_a=\breve{\bf e}_a$, i.\  e.\  we accept, that the
arbitrary tetrad ${\bf e}_a$ in the equations (2.1) coincides with
the proper tetrad of the spinor field $\psi $. In this case the scale
factors $h^i{}_a$ in the equations (2.1) coincide with coefficients
$\breve h^i{}_a$, determined by matrix (1.12).

The Dirac equations in this case are written as the equations on the Ricci
rotation coefficients  $\breve\Delta _{a,bc}$  and the invariants
of the spinor field \cite{7}
$$
\begin{array}{c}
 \breve D_a\ln\rho +\breve\Delta _{b,a}{}^b=2m\breve \sigma _a\sin\eta, \\
\breve D^a\eta +\displaystyle\frac 12\varepsilon ^{abcd}\breve\Delta _{b,cd}=
2m\breve \sigma ^a\cos \eta .
\end{array}
\eqno(2.3)
$$

Here $ \breve D_a =\breve h^i{}_a\partial _i=\{\pi ^i\partial _i,
\xi ^i\partial _i, \sigma ^i\partial _i, u^i\partial _i\}$;
$\breve\sigma _a=\breve\sigma ^a=(0,0,1,0) $ are the components of the
vector $\breve{\bf e}_3$ in the proper basis.

Coefficients $\breve\Delta _{a,bc}$ in the equations (2.3) are connected
with the scale factors $\breve h^i{}_a$ by relation
$$
\breve\Delta _{a,bc}=\frac 12\bigl[ \breve h^j{}_a(\breve D_b\breve h_{jc}
-\breve D_c\breve h_{jb})+\breve h^j{}_c(\breve D_a\breve h_{jb}+
\breve D_b\breve h_{ja})-\breve h^j{}_b(\breve D_a\breve h_{jc}+
\breve D_c\breve h_{ja})\bigr] .
\eqno(2.4)
$$

The equations (2.3) are the identical record of the Dirac equations
in the proper basis $\breve{\bf e}_a$. These equations can
be received also from the Dirac equations in (2.1), if to replace in them
derivatives $\nabla _i\psi $
via the formula (1.16) and to produce algebraic transformations.

The replacement in equations (2.3) coefficients $ \breve\Delta _{a,bc}$
via $\breve h^i{}_a$ yields the following system of
the invariant tensor equations \cite{7}
$$
\begin{array}{c}
\nabla _i\rho\pi ^i=0,\qquad \nabla _i\rho\xi ^i=0,\qquad
\nabla _i\rho\sigma ^i=2m\rho\sin\eta,\qquad \nabla _i\rho u^i=0, \\
\nabla ^i\eta -\displaystyle\frac 12 \varepsilon ^{ijms}\left(\pi _j
\nabla _m\pi _s+ \xi _j\nabla _m\xi _s+\sigma _j\nabla _m\sigma _s-
u_j\nabla _mu_s\right)= 2m\sigma ^i\cos\eta .
\end{array}
\eqno(2.5)
$$

The Einstein equations in the proper basis $\breve{\bf e}_a$ it is
convenient to write as
$$
\breve R_{ab}=\kappa \breve T_{ab}+
\left( \frac 12\kappa m\rho \cos\eta +\lambda \right) g_{ab}.
\eqno(2.6)
$$

For transformation of the Einstein equations to the form (2.6) it is
necessary to take into account, that in view of the equations (2.1)
the follow equation is valid
$$
R=-\kappa T_a{}^a+4\lambda =\kappa m\rho \cos \eta
+4\lambda .
\eqno(2.7)
$$

The expression of tetrad components of the energy--momentum tensor in
gauge ${\bf e}_a=\breve{\bf e}_a$ is obtained in \cite{7}:
$$
\breve T_{ab}=\frac 14\rho \left[ -\breve\sigma _b\breve D_a\eta -
\breve \sigma _a\breve D_b\eta +
\frac 12\breve\sigma _e\left( \breve\Delta _{a,cd}\varepsilon_b{}^{cde}+
\breve\Delta _{b,cd}\varepsilon _a{}^{cde}\right)\right].
\eqno(2.8)
$$

The tensor components $ \breve R_{ab} $  we can express via the Ricci
rotation coefficients
$$
\breve R_{ab}=\frac 1{\sqrt {-g}}\partial _j\left[\sqrt
{-g}\left( \breve h^j{}_c \breve \Delta _{b,a}{}^c-\breve h^j{}_b\breve
\Delta _{c,a}{}^c\right)\right]- \breve \Delta _{f,b}{}^c\breve \Delta
_{c,a}{}^f+\breve \Delta _{c,a}{}^c\breve \Delta _{f,b}{}^f.
\eqno(2.9)
$$
Here $g=\det\|g _{ij}\|$.

Let us give also the expression of the Ricci rotation coefficients
$\breve\Delta _{a,bc}$ immediately via the vector components
$ \pi _i $, $ \xi _i $, $ \sigma _i $, $u_i $, which is obtained from the
formula (2.4), if replace the components $\breve h^i{}_a$ by the formula
(1.12)
$$  \vspace{1\jot}
\begin{array}{rcl}
&&\breve\Delta _{1,12}
=\xi ^i\breve D_1\pi _i-\pi ^i\breve D_2\pi_i,\qquad\vspace{1\jot}
\breve\Delta _{2,12}
=-\pi ^i\breve D_2\xi _i+\xi ^i\breve D_1\xi_i,\\ \vspace{1\jot}
&&\breve\Delta _{3,12}
= \frac 12\bigl(\xi ^i\breve D_3\pi _i-\pi^i\breve D_3\xi _i\bigr)
+\frac 12\bigl( -\sigma ^i\breve D_2\pi _i-\pi ^i\breve D_2\sigma _i
+ \sigma ^i\breve D_1\xi _i+\xi ^i\breve D_1\sigma_i\bigr), \\ \vspace{1\jot}
&&\breve\Delta _{4,12}=
\frac 12\bigl(\xi ^i\breve D_4\pi _i-\pi^i\breve D_4\xi _i\bigr) +
\frac 12\bigl( -u^i\breve D_2\pi _i-\pi ^i\breve D_2u_i+ u^i\breve D_1\xi _i
+\xi ^i\breve D_1u_i\bigr), \\ \vspace{1\jot}
&&\breve\Delta_{1,23}=
\frac 12\bigl( \sigma ^i\breve D_1\xi _i-\xi ^i\breve D_1\sigma_i\bigr) +
\frac 12\bigl( \pi ^i\breve D_2\sigma _i+\sigma ^i\breve D_2\pi_i-
\pi ^i\breve D_3\xi _i-\xi ^i\breve D_3\pi _i\bigr), \\ \vspace{1\jot}
&&\breve\Delta_{2,23}=
\sigma ^i\breve D_2\xi _i-\xi ^i\breve D_3\xi _i,\qquad \breve\Delta
_{3,23}= -\xi ^i\breve D_3\sigma _i+\sigma ^i\breve D_2\sigma _i,\\ \vspace{1\jot}
&&\breve\Delta _{4,23}=
\frac 12\bigl( \sigma ^i\breve D_4\xi _i-\xi ^i\breve D_4\sigma _i\bigr) +
\frac 12\bigl( \sigma ^i\breve D_2u_i+u^i\breve D_2\sigma _i-
\xi ^i\breve D_3u_i-u^i\breve D_3\xi _i\bigr), \\ \vspace{1\jot}
&&\breve\Delta _{1,31}=
-\sigma ^i\breve D_1\pi _i+\pi ^i\breve D_3\pi _i,\qquad
\breve\Delta _{3,31}=
\pi ^i\breve D_3\sigma _i-\sigma ^i\breve D_1\sigma _i,\\ \vspace{1\jot}
&&\breve\Delta _{2,31}=
\frac 12\bigl( \pi ^i\breve D_2\sigma _i-\sigma ^i\breve D_2\pi _i\bigr)
+\frac 12\bigl( \pi ^i\breve D_3\xi _i+\xi ^i\breve D_3\pi _i
-\xi ^i\breve D_1\sigma _i-\sigma ^i\breve D_1\xi _i\bigr), \\ \vspace{1\jot}
&&\breve\Delta _{4,31}=
\frac 12\bigl( \pi ^i\breve D_4\sigma _i-\sigma ^i\breve D_4\pi _i\bigr)
+\frac 12\bigl( \pi ^i\breve D_3u_i+u^i\breve D_3\pi _i-
\sigma ^i\breve D_1u_i-u^i\breve D_1\sigma _i\bigr), \\ \vspace{1\jot}
&&\breve\Delta _{1,14}=u^i\breve D_1\pi _i-\pi ^i\breve D_4\pi _i,\qquad
\breve\Delta _{4,14}=-\pi ^i\breve D_4u_i+u^i\breve D_1u_i,\\ \vspace{1\jot}
&&\breve\Delta _{2,14}=
\frac 12\bigl( u^i\breve D_2\pi _i-\pi ^i\breve D_2u_i\bigr)
+\frac 12\bigl( \xi ^i\breve D_1u_i+u^i\breve D_1\xi _i-
\pi ^i\breve D_4\xi _i-\xi ^i\breve D_4\pi _i\bigr), \\ \vspace{1\jot}
&&\breve\Delta _{3,14}=
\frac 12\bigl( u^i\breve D_3\pi _i-\pi ^i\breve D_3u_i\bigr)
+\frac 12\bigl( \sigma ^i\breve D_1u_i+u^i\breve D_1\sigma _i-
\pi ^i\breve D_4\sigma _i-\sigma ^i\breve D_4\pi _i\bigr), \\ \vspace{1\jot}
&&\breve\Delta _{1,24}=
\frac 12\bigl( u^i\breve D_1\xi _i-\xi ^i\breve D_1u_i\bigr)
+\frac 12\bigl( \pi^i\breve D_2u_i+u^i\breve D_2\pi _i-
\pi ^i\breve D_4\xi _i-\xi ^i\breve D_4\pi _i\bigr), \\ \vspace{1\jot}
&&\breve\Delta _{2,24}=
u^i\breve D_2\xi _i-\xi ^i\breve D_4\xi _i,\qquad
\breve\Delta _{4,24}=-\xi ^i\breve D_4u_i+u^i\breve D_2u_i,\\ \vspace{1\jot}
&&\breve\Delta _{3,24}=
\frac 12\bigl( u^i\breve D_3\xi _i-\xi ^i\breve D_3u_i\bigr)
+\frac 12\bigl(\sigma ^i\breve D_2u_i+u^i\breve D_2\sigma _i-
\xi ^i\breve D_4\sigma _i-\sigma ^i\breve D_4\xi _i\bigr), \\ \vspace{1\jot}
&&\breve\Delta _{1,34}=
\frac 12\bigl( u^i\breve D_1\sigma _i-\sigma ^i\breve D_1u_i\bigr)
+\frac 12\bigl( \pi ^i\breve D_3u_i+u^i\breve D_3\pi _i-
\pi ^i\breve D_4\sigma _i-\sigma ^i\breve D_4\pi _i\bigr), \\ \vspace{1\jot}
&&\breve\Delta _{2,34}=
\frac 12\bigl( u^i\breve D_2\sigma _i-\sigma ^i\breve D_2u_i\bigr)
+\frac 12\bigl( \xi ^i\breve D_3u_i+u^i\breve D_3\xi _i-
\xi ^i\breve D_4\sigma _i-\sigma ^i\breve D_4\xi _i\bigr), \\ \vspace{1\jot}
&&\breve\Delta _{3,34}=
u^i\breve D_3\sigma _i-\sigma ^i\breve D_4\sigma _i,\qquad
\breve\Delta _{4,34}=-\sigma ^i\breve D_4u_i+u^i\breve D_3u_i.
\end{array}
\eqno(2.10)  $$

At an integration of the Einstein--Dirac equations it is useful to
take into account also the matrix of the components of
the energy--momentum tensor,
which is obtained in correspondence with definition (2.8):
$$
\breve T_{ab}=
\frac 14\rho \left\| \begin{array}{cccc}
 2\breve\Delta _{1,24}
& \breve\Delta _{2,24}-\breve\Delta _{1,14}
& -\breve D_1\eta +\breve\Delta _{3,24} & \breve\Delta _{4,24}-
\breve\Delta _{1,12} \\
\breve\Delta _{2,24}-\breve\Delta _{1,14}
& -2\breve\Delta _{2,14} & -\breve D_2\eta -\breve\Delta _{3,14}
& -\breve\Delta _{2,12}-\breve\Delta _{4,14}            \\
-\breve D_1\eta +\breve\Delta _{3,24}
& -\breve D_2\eta -\breve\Delta _{3,14}
& -2\breve D_3\eta
& -\breve D_4\eta -\breve\Delta _{3,12}                  \\
\breve\Delta _{4,24}-\breve\Delta _{1,12}
& -\breve\Delta _{2,12}-\breve\Delta _{4,14}
& -\breve D_4\eta -\breve\Delta _{3,12}
& -2\breve\Delta _{4,12}
			    \end{array}\right\| .
\eqno(2.11)
$$

The equations (2.3), (2.6), (2.9) --- (2.11) in the given system of
coordinate $x^i $ represent a closed set of equations for determining
the functions $ \pi _i(x^j) $, $ \xi _i(x^j) $, $ \sigma _i(x^j) $,
$ u_i(x^j) $, $ \rho (x^j) $, $ \eta (x^j) $.
The contravariant components of the metric tensor of the Riemannian space
are expressed via $\pi _i(x^j) $, $\xi _i(x^j) $, $\sigma _i(x^j) $,
$u_i(x^j) $ by the relation
$$
g^{ij}=\breve h^i{}_a\breve h^j{}_bg^{ab}=
\pi ^i\pi ^j+\xi ^i\xi ^j+\sigma ^i\sigma ^j-u^iu^j.
\eqno(2.12)
$$

  \section{
General exact solution of the Einstein--Dirac equations
in homogeneous space
  }

Let us consider a four-dimensional Riemannian space referred to
synchronous system of coordinates with variables $x^i $, in which
on definition are
valid the equations
$$
g_{44}=g^{44}=-1, \qquad
g_{4\alpha }=g^{4\alpha }=0, \qquad \alpha =1,2,3.
\eqno(3.1)
$$
We shall seek the solutions of the equations (2.3), (2.5), (2.6),
(2.9) --- (2.11) in the synchronous system of coordinate in the
supposition, that all unknown functions depend only from parameter
$x^4=t $. Thus, it is suposed, that the space $V$ is the homogeneous
Bianchi 1 type Riemannian space.

In this case the equations (2.5) are written as
$$
\begin{array}{c}
\partial _4(\sqrt{-g}\,\rho \pi ^4)=0, \qquad
\partial _4(\sqrt{-g}\,\rho \xi ^4)=0,
\qquad \partial _4(\sqrt{-g}\,\rho u^4)=0, \vspace{1\jot} \\
\partial _4(\sqrt{-g}\,\rho \sigma ^4)=2m\sqrt{-g}\,\rho\sin\eta,\qquad
\partial _4\eta =-2m\sigma ^4\cos \eta .
\end{array}
\eqno(3.2)
$$

The set of equations (3.2) and equation
$$
g^{44}\equiv\pi ^4\pi ^4+\xi ^4\xi ^4+\sigma ^4\sigma ^4-u^4u^4=-1
\eqno(3.3)
$$
are the complete set for determining the quantities $\pi ^4$, $\xi ^4$,
$ \sigma ^4 $, $u^4 $, $ \eta $, $ \rho\sqrt{-g} $.
The general solution of the equations (3.2), (3.3) has the form
$$
\begin{array}{c}
\displaystyle\frac {C_\rho }{\rho \sqrt{-g}}
=\frac {\pi ^4}{C_{\pi} }
=\frac {\xi ^4}{C_{\xi }}=\frac {u^4}{C_u}
=\frac 1{\sqrt{1+C_\sigma ^2 \cos ^2(2mt+\varphi )}},   \\
\sigma ^4=\displaystyle\frac {\varepsilon C_\sigma \sin (2mt+\varphi )}
{\sqrt {1+ C_\sigma ^2\cos ^2(2mt+\varphi )}}, \qquad
\exp i\eta=\varepsilon \frac {1+ iC_\sigma\cos (2mt+\varphi )}
{\sqrt{1+C_\sigma ^2\cos ^2(2mt+\varphi )}},
\end{array}
\eqno(3.4)
$$
where $\varphi $, $C_\pi$, $C_\xi$, $C_\sigma $, $C_u\geq 1$, $C_\rho >0$
are the integration constants; coefficient $ \varepsilon $ can take
any of the two values $+1$ or $-1$. By virtue of the condition of
synchronism (3.3) the constans $C $ are connected by the relation
$$
(C_\pi )^2+(C_\xi )^2+(C_\sigma)^2-(C_u)^2=-1.
\eqno(3.5)
$$

Let us introduce the notation
$h_a=\breve h^4{}_a=(\pi ^4, \xi ^4, \sigma ^4, u^4)$.
From the solution (3.4) we find
$$
\{ h_1, h_2, h_4 \}
=\frac 1{\sqrt{1+C_\sigma ^2\cos ^2(2mt +\varphi)}}\{C_\pi, C_\xi, C_u\}.
\eqno(3.6)
$$
It is clear from here, that by virtue of solution (3.4) the directions
of the three-dimensional vector with components $h_1$, $h_2$, $h_4$
does not depend on parameter $t$.

From definition (2.4) it follows, that the Ricci rotation coefficients
$\breve\Delta _{a,bc}$ can be presented in the form
$$
\breve\Delta _{a,bc}=\frac 12(h_bs_{ac}-h_cs_{ab}-h_aa_{bc}),
\eqno(3.7)
$$
where by definition
$$
\begin{array}{c}
s_{ab}=s_{ba}=\breve h^i{}_a\partial _4\breve h_{ib}+
\breve h^i{}_b\partial _4\breve h_{ia}, \\
a_{ab}=-a_{ba}=\breve h^i{}_a\partial _4\breve h_{ib}-
\breve h^i{}_b\partial _4\breve h_{ia}.
\end{array}
\eqno(3.8)
$$
The quantities $h_a $ in the formula (3.7) are determined
by solution (3.4) as the functions of parameter $t $, therefore the
relation (3.8) expresses 24 dependent functions $ \breve\Delta _{a,bc}$
only through 16 functions $s_{ab}$, $a_{ab}$.

By virtue of definition (3.8) the quantities $s_{ab}$ in a synchronous
system of the coordinates satisfy  the identities
$$
\begin{array}{c}
h^bs_{ab}=0, \qquad
s_a{}^a=2\partial _4\ln \sqrt {-g}.
\end{array}
\eqno(3.9)
$$

From the first identity in (3.9) it follows, that the ten symmetrical
components $s_{ab}$ contains generally the six independent ones.

From the equations (2.3) it follows, that
the antisymmetric quantities $a_{ab}$ are determined by equality
$$
a_{ab}=4m\left[ (\breve \sigma _ah_b-\breve \sigma _bh_a)\sin \eta -
\varepsilon _{abcd}\breve \sigma ^ch^d\cos \eta \right].
\eqno(3.10)
$$

Using the equations (3.9), (3.10)
and definitions (2.9), (2.11) for $T_{ab}$, $R_{ab}$ we can transform
the Einstein equations (2.6) into the equivalent
set of equations
$$
\begin{array}{c}
\partial _4(\sqrt{-g}\,s_{ab})-2\sqrt{-g}\,\left( m\cos\eta
+\frac 18\kappa\rho\right)\left(\varepsilon _{cefa}s_b{}^f+
\varepsilon _{cefb}s_a{}^f\right)h^c\breve \sigma ^e-   \vspace{1\jot}\\
-2m\sqrt{-g}\,\bigl( h_as_{bc}+h_bs_{ac}\bigr)\breve \sigma ^c\sin \eta
=\bigl(\kappa m\rho\sqrt{-g}\cos\eta +2\lambda\sqrt{-g}\bigr)\bigl(
g_{ab}+h_ah_b\bigr),             \vspace{1\jot}\\
\left( s_a{}^a\right)^2-s_{ab}s^{ab}=8(\kappa m\rho\cos\eta +\lambda ).
\end{array}
\eqno(3.11)
$$

The quantity $\rho\sqrt{-g}\cos\eta$ in the right part of
the first equation (3.11) by virtue of the solution (3.4) is constant
$$
\rho\sqrt{-g}\cos\eta =\varepsilon C_\rho.
\eqno(3.12)
$$
The first equation in (3.11) is obtained by contracting of the Einstein
equations in (2.6) with the tensor components $ \delta _c^a+h_ch^a $
by the index $a $. The second equation in (3.11) is obtained by
contracting of the Einstein equations in (2.6) with the components
of tensor $g ^ {ab} +2h^ah^b $ by the indices $a$, $b$.

The contracting of the equations (3.11) with $g^{ab}$ by the indices $a$,
$b$ gives the equation
$$
\partial _4\partial _4\sqrt{-g}=\frac 32\kappa m\varepsilon C_\rho
+3\lambda \sqrt{-g}
\eqno(3.13)
$$
for definition of the quantity $\sqrt{-g}$.

If the cosmological constant $\lambda$ is positive $\lambda >0 $,
the solution of the equation (3.13) has the form
$$
\sqrt{-g}=\frac {\varepsilon \kappa m}{2\lambda }C_\rho \left[ -1+
f_1\sinh \bigl(\sqrt{3\lambda }\,t\bigr) +
f_2\cosh \bigl(\sqrt{3\lambda }\,t\bigr)\right],
\eqno(3.14)
$$
where $f_1 $, $f_2 $ are arbitrary constants.

If $\lambda <0$, for $\sqrt{-g}$ is obtained
$$
\sqrt{-g}=\frac {\varepsilon \kappa m}{2\lambda }C_\rho \left\{
-1+f\sin\bigl[\sqrt{-3\lambda }\, (t-t_\circ )\bigr]\right\},
\eqno(3.15)
$$
where $f $, $t_0 $ are the integration constants.

The case $\lambda =0$ has been considered in \cite{14,15}.

As a direction of the three-dimensional vector with components
$h_1 $, $h_2 $, $h_4 $ does not depend on parameter $t $, then by the
constant Lorentz transformation of the basis vectors
$\breve{\bf e}_1$, $\breve {\bf e}_2$,
$\breve{\bf e}_4$ it is
possible to transform the components $h_1$, $h_2$ to zero. The initial set
of equations (2.3), (2.6), (2.9) -- (2.11) is invariant under the
arbitrary constant Lorentz transformation of the vectors
$\breve{\bf e}_1$, $\breve{\bf e}_2$,
$\breve{\bf e}_4$. Therefore without loss of generallity
it is enough to consider a solution of the equations
(2.3), (2.6), (2.9) - (2.11) only at $h_1=h_2=0 $. Under
this condition the first equation in (3.11) can be written in the form

$$ 
\begin{array}{rcl}
&&\hphantom{.......}\partial _4(s_{33}\sqrt{-g}\, )-
4m\sigma ^4\sin \eta \, s_{33}\sqrt{-g}=
\bigl(\varepsilon\kappa mC_\rho +2\lambda\sqrt{-g}\,\bigr) (u^4)^2, \\
&&\partial _4(s_{23}\sqrt{-g}\, )-2m\sigma ^4\sin \eta \, s_{23}\sqrt{-g}-
\biggl( 2m\cos\eta +\frac 14\kappa\rho\biggr) u^4s_{13}\sqrt{-g}=0,\\
&&\partial _4(s_{13}\sqrt{-g}\, )-2m\sigma ^4\sin\eta\, s_{13}\sqrt{-g}+
\biggl( 2m\cos\eta +\frac 14\kappa\rho\biggr)u^4s_{23}\sqrt{-g}=0, \\
&&\hphantom{.......}\partial _4(s_{11}\sqrt{-g}\, )+
2\biggl(2m\cos \eta +\frac 14\kappa\rho \biggr)u^4s_{12} \sqrt{-g}=
 \varepsilon \kappa mC_\rho +2\lambda \sqrt{-g},    \\
&&\hphantom{.......}\partial _4(s_{22}\sqrt{-g}\, )
   -2\biggl( 2m\cos \eta +\frac 14\kappa \rho \biggr)u^4s_{12}
   \sqrt{-g}=\varepsilon \kappa mC_\rho +2\lambda \sqrt{-g},     \\
&&\hphantom{.......}\partial _4(s_{12}\sqrt{-g}\, )-\biggl(2m\cos\eta +
   \frac 14\kappa \rho \biggr)u^4(s_{11}-s_{22})\sqrt{-g}=0,      \\
&&\partial _4(s_{14}\sqrt{-g}\, )+\biggl(2m\cos\eta +\frac 14\kappa\rho
   \biggr)u^4s_{24} \sqrt{-g}-2mu^4\sin\eta\, s_{31}\sqrt{-g}=0,  \\
&&\partial _4(s_{24}\sqrt{-g}\, )-\biggl(2m\cos\eta +\frac 14\kappa\rho
   \biggr)u^4s_{14} \sqrt{-g}-2mu^4\sin\eta\, s_{23}\sqrt{-g}=0, \\
\vspace{1.1\jot}
&&\partial _4(s_{34}\sqrt{-g}\, )-2m\sigma ^4\sin\eta\,
s_{34}\sqrt{-g}=u^4(\varepsilon\kappa mC_\rho\sigma ^4 +2m\sin\eta\,
   s_{33}\sqrt{-g}\, ),  \\
\vspace{2\jot }
&&\partial _4(s_{44}\sqrt{-g}\, )-4mu^4\sin\eta\, s_{34}\sqrt{-g}=
\bigl(\varepsilon\kappa mC_\rho +
2\lambda\sqrt{-g}\,\bigr) (\sigma ^4)^2.
\end{array}
\eqno(3.16)
$$

 The general solution for $s_{ab} $ has the form:
$$
\begin{array}{l}
s_{11}=\rho u^4\left[ -
\displaystyle\frac 13N+\frac 2{3C_\rho C_u}\partial _4
\sqrt{-g}+\displaystyle\frac 12B\sin 2(\zeta +\beta )\right],  \\
s_{22}=\rho u^4\left[ -\displaystyle\frac 13N+\frac 2{3C_\rho C_u}\partial _4
\sqrt{-g}-\displaystyle\frac 12B\sin 2(\zeta +\beta )\right],\\
s_{33}=\displaystyle\frac 23\rho (u^4)^3\biggl(
N+\displaystyle\frac 1{C_\rho C_u}\partial _4\sqrt{-g}\biggr), \\
s_{44}=\displaystyle\frac 23\rho u^4(\sigma ^4)^2\biggl(
N+\displaystyle\frac 1{C_\rho C_u}\partial _4\sqrt{-g}\biggr), \\
s_{12}=-\displaystyle\frac 12\rho u^4B\cos 2(\zeta +\beta ),\\
s_{34}=\displaystyle\frac 23\rho \sigma ^4(u^4)^2\biggl(
N+\displaystyle\frac 1{C_\rho C_u}\partial _4\sqrt{-g}\biggr), \\
s_{13}=\displaystyle\frac 12\rho (u^4)^2A\cos (\zeta +\alpha ), \quad
s_{14}=\displaystyle\frac 12\rho u^4\sigma ^4A\cos (\zeta +\alpha ),\\
s_{23}=\displaystyle\frac 12\rho (u^4)^2A\sin (\zeta +\alpha ), \quad
s_{24}=\displaystyle\frac 12\rho u^4\sigma ^4A\sin (\zeta +\alpha ),
\end{array}
\eqno(3.17)
$$
where $A$, $B$, $N$, $\alpha $, $\beta $ are the arbitrary constants;
the quantity $ \sqrt{-g} $ in the equations (3.17) is defined by the
equalities (3.14), (3.15); the quantity $\zeta $ is defined by the
follow relation
$$
\zeta =\int \left( 2m\cos \eta +\frac
14\kappa \rho \right)u^4\, dt=\varepsilon \mathop{\rm arctg}\left(
\frac{\tan  (2mt+\varphi )} {\sqrt{1+C_\sigma ^2}}\right) +
\frac 14\kappa \tau .
\eqno(3.18)
$$
The integral $\tau =\int\rho u^4dt$ in the formula
(3.18) depends from the value of cosmologic constant $\lambda $
and will be calculated further.

 Substituting in the second equation in (3.11) the components $s_{ab}$
by the formulas (3.17), we obtain a connection between the integration
constants $A $, $B $, $N $ and $f_1 $, $f_2 $. In the case $\lambda >0 $
is obtained the following relation
$$
f_2^2-f_1^2=1-\frac {\lambda }{\kappa ^2m^2}C_u^2\left( \frac
14A^2+\frac 14B^2+\frac 13N^2\right)  \leq 1.
$$

 It is obvious, that if $f_2^2\geq f_1^2 $, then the formula (3.14) for
$\sqrt{-g}$ can be presented as
$$
\sqrt{-g}=\frac {\varepsilon\kappa m}{2\lambda }C_\rho \left\{ -1 +
f\cosh \bigl[\sqrt{3\lambda}\, (t-t_\circ )\bigr]\right\}, \qquad
|f|\leq 1,
\eqno(3.19 a)
$$
 where $t_\circ $, $f $ are the arbitrary constants, $f^2=f_2^2-f_1^2 $.

 The singular points of the solution (3.19 a) are determined by the equation
$f\cosh \bigl[\sqrt{3\lambda }\, (t-t_\circ )\bigr]=1$.
The solution has one singular point, if $f=1 $ and two singular points,
if $0<f<1$.
If $f\leq 0$ (in this case $\varepsilon =-1$), the solution has no
singular points.

 If $f_2^2<f_1^2$, for $\sqrt{-g}$ the equality is valid
$$
\sqrt{-g}=\frac {\varepsilon\kappa m}{2\lambda }C_\rho \left\{ -1 +
f\sinh \bigl[\sqrt{3\lambda }\, (t-t_\circ )\bigr]\right\},
\eqno(3.19 b)
$$
where $f^2=f_1^2-f_2^2$. In this case the solution has one singular point,
defined by the equation
$f\sinh \bigl[\sqrt{3\lambda }\, (t-t_\circ )\bigr]=1$.

 If $\lambda <0$, the following formula for the integration constant
$f$ in (3.15) is obtained
 $$
 f^2=1-\frac \lambda {\kappa ^2m^2} C_u^2\left (\frac 14A^2 +
 \frac 14B^2 +\frac 13N^2\right) \geq 1.
 $$
 In this case there are indefinitely many singular points, are determined
by the equation $\sqrt{-g}=0$.

 The condition $\sqrt{-g}>0$ yields a restriction on the values of the
integration constants and the domain of existence of the solution.

Using definition (1.14) and formulas (3.4), we can write out solution for
 the components of spinor field in the proper basis
$$ 
\breve \psi =\pm   \left\| \begin{array}{c}
 0   \\
i\sqrt{\varepsilon C_\rho \displaystyle\frac {1+
iC_\sigma \cos(2mt+\varphi )}{2\sqrt{-g}}}   \\
         	  0 \\
	i\sqrt{\varepsilon C_\rho \displaystyle\frac {1-iC_\sigma \cos(2mt+
                   \varphi )}{2\sqrt{-g}}}
\end{array}\right\|,
$$
where $\sqrt{-g}$, depending on a sign $ \lambda $, is defined by the
formula (3.15) or formulas (3.19 a,b).

The equations (3.4), (3.10), (3.17) completely determine the Ricci
rotation coefficients $\breve\Delta _{a,bc} $ by the formula (3.7)
and represent the first integral of the equations (2.3), (2.5), (2.6),
(2.9) --- (2.11). For determining of the general solution of the
equations (2.3), (2.5), (2.6), (2.9) --- (2.11) it is now enough to
integrate the equations (3.8), from which follows
$$
\partial _4\breve h_{ib}=\frac 12\bigl( s_{ab}+a_{ab}\bigr)\breve h_i{}^a.
\eqno(3.20)
$$
In view of definition (1.12) the quantities $\breve h_i{}^a$ and found
solutions (3.10), (3.17) for $s_{ab}$, $a_{ab}$, the equation (3.20) is
possible to transform to the form
$$\begin{array}{rcl}
&&\displaystyle\frac{d{}}{d\tau }\left( u^4\sigma _j-\sigma ^4u_j\right)=\frac 14A
\bigl[ \pi _j\cos (\zeta +\alpha )+\xi _j\sin (\zeta +\alpha )\bigr], \\
&&\hphantom {....................}
+\left( u^4\sigma _j-\sigma ^4u_j\right)\left(
\displaystyle\frac 1{3\sqrt{-g}}\frac {d{}}{d\tau }\sqrt{-g}+
\frac 13N\right),           \\
&&\displaystyle\frac{d{}}{d\tau }\bigl( \xi _j+i\pi _j\bigr)
=(\xi _j+i\pi _j)\left( \frac 1{3\sqrt{-g}}\frac {d{}}{d\tau }\sqrt{-g}
-\frac 16N -i\frac {2m}\rho \cos \eta \right),   \vspace{1\jot}    \\
&&\hphantom {......}
-\displaystyle\frac i4\bigl( \xi _j-i\pi _j\bigr) B\exp \bigl[ -2i(\zeta
+\beta )\bigr] +\frac i4A\bigl( u^4\sigma _j
-\sigma ^4u_j\bigr) \exp [-i(\zeta +\alpha )].
\end{array}
\eqno(3.21)
$$

 The differential $d\tau $ in (3.21) is defined by equality
 $d\tau =\rho u^4dt$.
 The set of equations (3.21) should be supplemented by the condition of
synchronism
$$
g_{4\alpha }\equiv \sigma _4\sigma _\alpha -u_4u_\alpha =0,
\qquad \alpha =1,2,3.
\eqno(3.22)
$$

 For the parameter $ \tau $ we have
$$
\tau =\int \rho u^4dt=C_\rho C_u\int \frac {dt}{\sqrt{-g}}.
$$
If the cosmological constant is positive $\lambda >0$ and
$\kappa m\ne 0$, a calculation by means of the equations (3.19 a,b)
gives
$$
\begin{array}{rcl}
\tau &=&\displaystyle\frac {2\lambda C_u}{\varepsilon \kappa m}\int
\frac{dt}{-1+ f\sinh \bigl[\sqrt{3\lambda }\, (t-t_\circ )\bigr] }   \\
&=&\displaystyle\frac {2\sqrt{\lambda }\, C_u}{\sqrt{3}\,\varepsilon \kappa m}
\frac 1{\sqrt{1+f^2}}
\ln\left| \frac{f-\sqrt{f^2+1}+\tanh\frac 12\sqrt{3\lambda }\,
(t-t_0)}
{f+\sqrt{f^2+1}+\tanh\frac 12\sqrt{3\lambda }\,
(t-t_0)}\right| +\tau _0
\end{array}
\eqno(3.23 a)
$$
 and
$$
\begin{array}{rcl}
\tau &=&\displaystyle\frac {2\lambda C_u}{\varepsilon \kappa m}\int
\frac {dt}{-1+f\cosh \bigl[\sqrt{3\lambda }\, (t-t_\circ )\bigr]}   \\
&=&\displaystyle\frac {2\sqrt{\lambda }\, C_u}{\sqrt{3}\,\varepsilon \kappa m}
\frac 1{\sqrt{1-f^2}}\ln\left|
\frac{f-1+\sqrt{1-f^2}\tanh {\frac 12}\sqrt{3\lambda }\,
(t-t_0)}
{f-1-\sqrt{1-f^2}\tanh \frac 12\sqrt{3\lambda }\,
(t-t_0)}\right|+\tau _0,
\end{array}
\eqno(3.23 b)
$$
where $\tau _0$ is the arbitrary constant.

 At $\lambda <0$  $(f^2>1)$ by means of (3.15) we obtain
$$
\begin{array}{rcl}
\tau &=&
\displaystyle\frac {2\lambda C_u}{\varepsilon \kappa m}\int \frac {dt}{-1+
f\sin\bigl[\sqrt{-3\lambda }\, (t-t_\circ )\bigr] }         \\
&=&\displaystyle\frac {2\sqrt{-\lambda }\, C_u}{\sqrt{3}\,\varepsilon \kappa m}
\frac 1{\sqrt{f^2-1}}\ln\left| \frac {\tan \frac 12\sqrt{
-3\lambda }\, (t-t_0)-f-\sqrt{f^2-1}}
{\tan \frac 12\sqrt{-3\lambda }\,
(t-t_0)-f+\sqrt{f^2-1}}\right|+\tau _0.
\end{array}
\eqno(3.24)
$$

Let us change now the unknowns functions
$(\pi _\lambda ,\xi _\lambda ,\sigma _\lambda ,u_\lambda )\to
(\pi _\lambda ^0,\xi _\lambda ^0,\theta _\lambda )$ in the equations
(3.21)
$$
\begin{array}{rcl}
\xi_{\lambda }+i\pi_{\lambda } &=&
\left( \xi_{\lambda }^0+i\pi_{\lambda }^0\right)
(\sqrt{-g}\, )^{1/3}\exp \biggl( -\frac16N\tau -i\zeta \biggr),   \\
\sigma_\lambda &=&
\theta _\lambda (\sqrt{-g}\, )^{1/3}u^4\exp\biggl( -\frac16N\tau\biggr),\\
u_\lambda &=&\theta _\lambda (\sqrt{-g}\, )^{1/3}\sigma ^4
\exp\biggl( -\frac 16N\tau\biggr).
\end{array}
\eqno(3.25)
$$
Here $ \zeta $ is defined by the relation (3.18), quantities $\sigma ^4$,
$u^4$ are defined by the solution (3.4), $\sqrt {-g}$ is determined by
equalities (3.15) or (3.19 a, b).
In outcome the condition of synchronism (3.22) is satisfied
identically, and the set of equations (3.21) goes over into the set of
linear equations with the constant coefficients:
$$
\begin{array}{c}
\displaystyle\frac{d{}}{d\tau }\bigl( \xi _\lambda ^0+i\pi _\lambda ^0\bigl)
=\frac i4\kappa \bigl( \xi _\lambda ^0+i\pi _\lambda ^0\bigl)
-\frac i4B\bigl( \xi _\lambda ^0-i\pi _\lambda ^0\bigl) e^{-2i\beta }
+\frac i4A\theta _\lambda e^{-i\alpha }, \\
\displaystyle\frac{d{}}{d\tau }\theta _\lambda
=\frac 14A\bigl(\pi _\lambda ^0\cos\alpha
+\xi _\lambda ^0\sin \alpha \bigl) +\frac 12N\theta _\lambda .
\end{array}
\eqno(3.26)
$$

At $j=4 $ the equations (3.21) are satisfied identically by virtue of
conditions $h_1=h_2=0 $.

Using definition (2.12), we obtain the expression for the spatial
components of metric tensor $g_{\alpha \beta}$ via
the vector components $\pi _\alpha ^0$, $\xi _\alpha ^0 $,
$\sigma _\alpha ^0$:
$$
g_{\alpha \beta }=\bigl(\sqrt{-g}\,\bigl) ^\frac 23 e^{-\frac 13 N\tau }
\bigl( \pi _\alpha ^0\pi _\beta ^0+
\xi _\alpha ^0\xi _\beta ^0+\theta _\alpha \theta _\beta \bigl) .
\eqno(3.27)
$$

To the equations (3.26) for each value of an index $ \lambda =1,2,3 $
corresponds the characteristic equation for an eigenvalue $q $
$$
\det \left\{\frac 14\left\| \begin{array}{ccc}
            B\sin 2\beta  & -B\cos 2\beta +\kappa & A\cos \alpha \\
    -B\cos 2\beta -\kappa & -B\sin 2\beta            & A\sin \alpha \\
            A\cos\alpha   &  A\sin\alpha             & 2N
                    \end{array}\right\|   -qI   \right\}=0,
$$
 which is possible to write as
 $$
 2 (N-2q) (16q ^2 +\kappa ^2-A^2-B^2) +
 A^2 [2N-B\sin 2 (\alpha -\beta)] =0.
 \eqno(3.28)
 $$

The solution of the equation (3.28) in general case is determined
by the Cardano formulas.
A simple solution of this equation is obtained in the case $A=0 $ and
$2N=B\sin 2(\alpha -\beta) $. A solution of equations (3.26) in the case
$A=0 $ has been considered in \cite{14,15}.

Let us consider the case, when is valid the equation
$2N=B\sin 2 (\alpha-\beta) $, $A\ne 0$. In this case for the eigenvalues
we have $q_1=\frac 12N $,
$q_2=\frac 14\bigl( A^2+B^2-\kappa ^2\bigr) ^{1/2}$
and $q_3=-\frac 14\bigl( A^2+B^2-\kappa ^2\bigr) ^{1/2}$.
If $0<A^2+B^2<\kappa ^2 $, it is possible to write a solution of the
equations (3.26) as follows
$$
\begin{array}{rcl}
\xi _\lambda ^0+i\pi _\lambda ^0&=
&e^{-i\alpha}\{-AQ_\lambda e^{\frac 12N\tau }
+[-(\kappa +Be^{2i(\alpha -\beta )})F_\lambda +
4i\Lambda G_\lambda ] \cos \Lambda \tau \\
&+&[-(\kappa +Be^{2i(\alpha -\beta )})G_\lambda -
4i\Lambda F_\lambda ]\sin \Lambda \tau \}, \\
\theta _\lambda &=&
[\kappa -B\cos 2(\alpha -\beta )]Q_\lambda e^{\frac 12N\tau }+
A(F_\lambda \cos \Lambda \tau +G_\lambda \sin \Lambda \tau ).
\end{array}
\eqno(3.29)
$$
 Here $ \Lambda = \frac 14\sqrt {\kappa ^2-A^2-B^2} $; $ F_\lambda $,
$G_\lambda $, $Q_\lambda $ are the integration constants connected
by the relation
$$
\varepsilon ^{\alpha \beta \lambda }F_\alpha G_\beta Q_\lambda =
(\kappa ^2-A^2-B^2)^{-3/2},
\eqno(3.30)
$$
where $\varepsilon ^{\alpha\beta\lambda}$ is the Levi-Chivita symbol.
A relation (3.30) expresses the equality between the value of
$\sqrt{-g}$, calculated from the solution (3.29), by its values (3.15),
(3.19 a,b).
For spatial components of the metric tensor by means of
the formulas (3.27) the following expression is obtained
$$
\begin{array}{rcl}
g_{\alpha \beta }&=&(\sqrt{-g}\, )^{2/3}e^{-\frac 13N\tau }
\Bigl\{
Q_\alpha Q_\beta \{ A^2+
[\kappa -B\cos 2(\alpha -\beta )]^2\} e^{N\tau } \\
&+&(F_\alpha G_\beta +
F_\beta G_\alpha )[M\sin 2\Lambda
\tau -4\Lambda B\sin 2(\alpha -\beta )\cos 2\Lambda \tau ] \\
&+&F_\alpha F_\beta [16\Lambda ^2+M+M\cos 2\Lambda \tau +
4\Lambda B\sin 2(\alpha -\beta )\sin 2\Lambda \tau ] \\
&+&G_\alpha G_\beta [16\Lambda ^2+M-M\cos 2\Lambda \tau -
4\Lambda B\sin 2(\alpha -\beta )\sin 2\Lambda \tau ] \\
&+&(F_\alpha Q_\beta +
F_\beta Q_\alpha )2\kappa Ae^{\frac 12N\tau }\cos \Lambda \tau \\
&+&(G_\alpha Q_\beta +G_\beta Q_\alpha )
2\kappa Ae^{\frac 12N\tau }\sin \Lambda \tau
\Bigr\},
\end{array}
\eqno(3.31)
$$
where the contracted notation is entered
$$
M=A^2+B^2 +\kappa B\cos 2 (\alpha -\beta).
$$

The quantity $\sqrt{-g}$ in the right part of equality (3.31) is defined
by the equations (3.19 a,b) or (3.15). It is obvious, that by the constant
transformation of the variable $x^1 $, $x^2 $, $x^3 $ it is always
possible to transform components
$F_\alpha $, $G_\alpha $, $Q_\alpha $ to the form
$$
\begin{array}{c}
F_\alpha =\bigl\{ (\kappa ^2-A^2-B^2)^{-1/2},0,0\bigr\}, \qquad
G_\alpha =\bigl\{0,(\kappa ^2-A^2-B^2)^{-1/2},0\bigr\}, \\
Q_\alpha =\bigl\{0,0,(\kappa ^2-A^2-B^2)^{-1/2}\bigr\}.
\end{array}
\eqno(3.32)
$$
The formula (3.31) at $A^2+B^2\ne 0 $ determine the oscillatory regime
of an motion to singular points of the solution.

If $A^2+B^2>\kappa ^2$, then the trigonometric functions in (3.29)
are replaced by the hyperbolic ones.

To the diagonal metric $g _ {ij} $ there corresponds the case, when
the integration constants $A $, $B $ in the equations (3.26) are equal to
zero. In this case we have
$$
\begin{array}{rcl}
\xi _\lambda ^0+i\pi _\lambda ^0&=&\kappa (-F_\lambda +
iG_\lambda ) \exp {\frac i4\kappa \tau },  \\
\theta _\lambda &=&
\kappa Q_\lambda \exp {\frac 12N\tau }.
\end{array}
\eqno(3.33)
$$

Taking into account equality (3.27), we obtain the solution for
the components of the metric tensor
$$
g_{\alpha \beta }=\kappa ^2\bigl( \sqrt{-g}\bigr)^{2/3}\left[
e^{-\frac 13N\tau }\bigl( F_\alpha F_\beta +G_\alpha G_\beta \bigr)
+e^{\frac 23N\tau }Q_\alpha Q_\beta \right].
\eqno(3.34)
$$
Here the quantity $\sqrt{-g}$ is defined by the formulas (3.15) or
(3.19 a,b), and $e^{N\tau}$ is defined by the relations
$$
\begin{array}{rcl}
&e^{N\tau }=\left(\displaystyle\frac{f-1+\sqrt{1-f^2}\tanh {\frac 12}
\sqrt{3\lambda }\, (t-t_0)}
{f-1-\sqrt{1-f^2}\tanh \frac 12\sqrt{3\lambda }\, (t-t_0)}
\right)^{2\varepsilon}   \qquad &\mbox{for}\quad
\lambda >0, \quad f_1^2<f_2^2,  \\
&e^{N\tau }=\left( \displaystyle\frac{f-\sqrt{f^2+1}+\tanh \frac
12\sqrt{3\lambda }\, (t-t_0)}{f+\sqrt{f^2+1}+\tanh \frac
12\sqrt{3\lambda }\, (t-t_0)}\right)^{2\varepsilon }\qquad &
\mbox{for}\quad \lambda >0, \quad f_1^2>f_2^2,  \\
&e^{N\tau }=\left( \displaystyle\frac
{\tan \frac 12\sqrt{-3\lambda }\, (t-t_0)-f-\sqrt{f^2-1}}
{\tan \frac 12\sqrt{-3\lambda }\, (t-t_0)-f+
\sqrt{f^2-1}}\right)^{2\varepsilon } \qquad &
\mbox{for}\quad \lambda <0.
\end{array}
\eqno(3.35)
$$

If $F_\lambda $, $G_\lambda $, $Q_\lambda $ to define by the relations
$$
F_\alpha =\left\{ \kappa ^{-1},0,0,\right\},\qquad
G_\alpha =\left\{ 0,\kappa ^{-1},0\right\},\qquad
Q_\alpha =\left\{ 0,0,\kappa ^{-1}\right\},
\eqno(3.36)
$$
then the metric (3.33) is diagonal.

If $A=B=N=0 $, then
$$
g_{\alpha \beta }=\kappa ^2\bigl( \sqrt{-g}\bigr)^{2/3}\left(
F_\alpha F_\beta +G_\alpha G_\beta +Q_\alpha Q_\beta \right),
\eqno(3.37)
$$
where $\sqrt{-g}$ is determined by the formulas (3.19 a), (3.15)
and $f^2=1$.

Some particular solutions of the Einstein--Dirac equations with the
cosmological constant in the homogeneous Riemannian space of
the Bianchi 1 type with the diagonal metric have been obtained in \cite{5,6}
(without cosmological constant see \cite{1,2,3,4,14,15}).

Let us consider the transformation of coordinates
$ (x ^\alpha, t) \to (x ^\alpha, \tau) $ defined by the equation
$ d\tau = \rho u^4dt $.
The function $\tau (t)$ is determined by relations (3.23 a,b), (3.24).
Quantities $\sqrt{-g}\, g^{4i}$ in the transformed system of coordinate
with variables $x ^\alpha $, $ \tau $ have the form
$$
(\sqrt{-g}\, g^{4\alpha})'=0,\qquad
 (\sqrt{-g}\, g^{44})'=\mbox {const}.
$$
From here follows, that in the system of coordinate with the variables
$x^\alpha $, $ \tau $ is satisfied the harmonicities condition
$$
\left[\partial _j(\sqrt{-g}\, g^{ij})\right]'=
\frac {d{}}{d\tau }(\sqrt{-g}\, g^{4i})'=0.
$$
Thus, parameter $ \tau $ is the time in the
harmonic system of coordinate.

\end{document}